\documentclass[iop]{emulateapj}
\usepackage{apjfonts}
\usepackage{verbatim}
\usepackage{xcolor}
\usepackage{natbib}
\usepackage{amssymb,amsmath,mathrsfs}
\tabletypesize{\scriptsize}

\usepackage[citecolor=blue,%
    filecolor=black,%
    linkcolor=blue,%
    pagebackref=true,
    colorlinks=true,%
    urlcolor=blue]{hyperref}

\def\h2{$\rm H_2$}

\def\MUV{$M_{\rm UV}$}

\newcommand{\msun}{M_{\odot}}

\begin{document}

\title{The Very Faint End of the UV Luminosity Function over Cosmic
  Time: Constraints from the Local Group Fossil Record}

\author{
Daniel R. Weisz\altaffilmark{1,2,4},
Benjamin D. Johnson\altaffilmark{1},
Charlie Conroy\altaffilmark{3,1}
}

\altaffiltext{1}{Department of Astronomy, University of California at Santa Cruz,
1156 High Street, Santa Cruz, CA, 95064; drw@ucsc.edu}
\altaffiltext{2}{Astronomy Department, Box 351580, University of Washington, Seattle, WA, USA}
\altaffiltext{3}{Harvard-Smithsonian Center for Astrophysics,
  60 Garden St., Cambridge MA 02138, USA}
\altaffiltext{4}{Hubble Fellow}


\begin{abstract}

  We present a new technique to estimate the evolution of the very
  faint end of the UV luminosity function (LF) out to $z\sim5$.
  Measured star formation histories (SFHs) from the fossil record of
  Local Group galaxies are used to reconstruct the LF down to
  \MUV$\sim-5$ at $z\sim5$ and \MUV$\sim-1.5$ at $z<1$.  Such faint
  limits are well beyond the current observational limits and are
  likely to remain beyond the limits of next generation facilities.
  The reconstructed LFs, when combined with direct measurements of the
  LFs at higher luminosity, are well-fit by a standard Schechter
  function with no evidence of a break to the faintest limits probed
  by this technique. The derived faint end slope, $\alpha$, steepens
  from $\approx-1.2$ at $z<1$ to $\approx-1.6$ at $4<z<5$.  We test
  the effects of burstiness in the SFHs and find the recovered LFs to
  be only modestly affected.  Incompleteness corrections for the
  faintest Local Group galaxies and the (unlikely) possibility of
  significant luminosity-dependent destruction of dwarf galaxies
  between high redshift and the present epoch are important
  uncertainties.  These and other uncertainties can be mitigated with
  more detailed modeling and future observations. The
  reconstructed faint end LF from the fossil record can therefore be
  a powerful and complementary probe of the high redshift faint
  galaxies believed to play a key role in the reionization of the
  Universe.

\end{abstract}

\keywords{galaxies: evolution --- galaxies: high-redshift ---
  Local Group --- galaxies: luminosity function, mass function}


\section{Introduction}

Knowledge of the faint end of the galaxy LF is essential for
understanding many aspects of the high redshift universe including the
extent to which faint galaxies contribute to the total cosmic SFR
density, to the reionization of the universe, and to gamma ray burst
rates.  The deepest {\it Hubble Space Telescope (HST)} images are
``only'' capable of reaching absolute magnitudes of
\MUV$\sim-17$,\footnote{Throughout this paper we assume AB magnitudes
  and use ``UV'' as shorthand for the GALEX FUV bandpass, which has a
  central wavelength of 1550\AA.}  at $z\sim5$ \citep{bouwens2014}.  However, several
phenomena require knowledge of the UV LF to much fainter limits.  For
example, models that are able to simultaneously reionize the universe with galaxies
by $z\sim6$ and match other constrains such as the Thomson optical
depth inferred from the cosmic microwave background must extrapolate
the UV LF to \MUV $\gtrsim -10$ \citep[for nominal assumptions
regarding the escape fraction of ionizing photons, the topology of the
IGM, etc.; see, e.g.,][]{dijkstra2004, kuhlen2012a, robertson2013,
  boylankolchin2014}.  Even the {\it James Webb Space Telescope
  (JWST)} may not reach such faint limits \citep{Windhorst2006},
suggesting that direct detection of the faint galaxies thought to be
key for reionization will prove elusive for decades.

Here, we present a new technique capable of probing the rest-frame UV
properties of very faint galaxies across cosmic time. Using star
formation histories (SFHs) of Local Group (LG) dwarf galaxies measured
from the analysis of deep color magnitude diagrams (CMDs), we are able
to synthesize the UV luminosity evolution of faint galaxies to high
redshifts. With this approach we can estimate the UV LF to limiting
magnitudes of \MUV$\sim-5$ at $z\sim5$ and $-1.5$ at $z<1$.

\begin{figure}[t!] 
\center \plotone{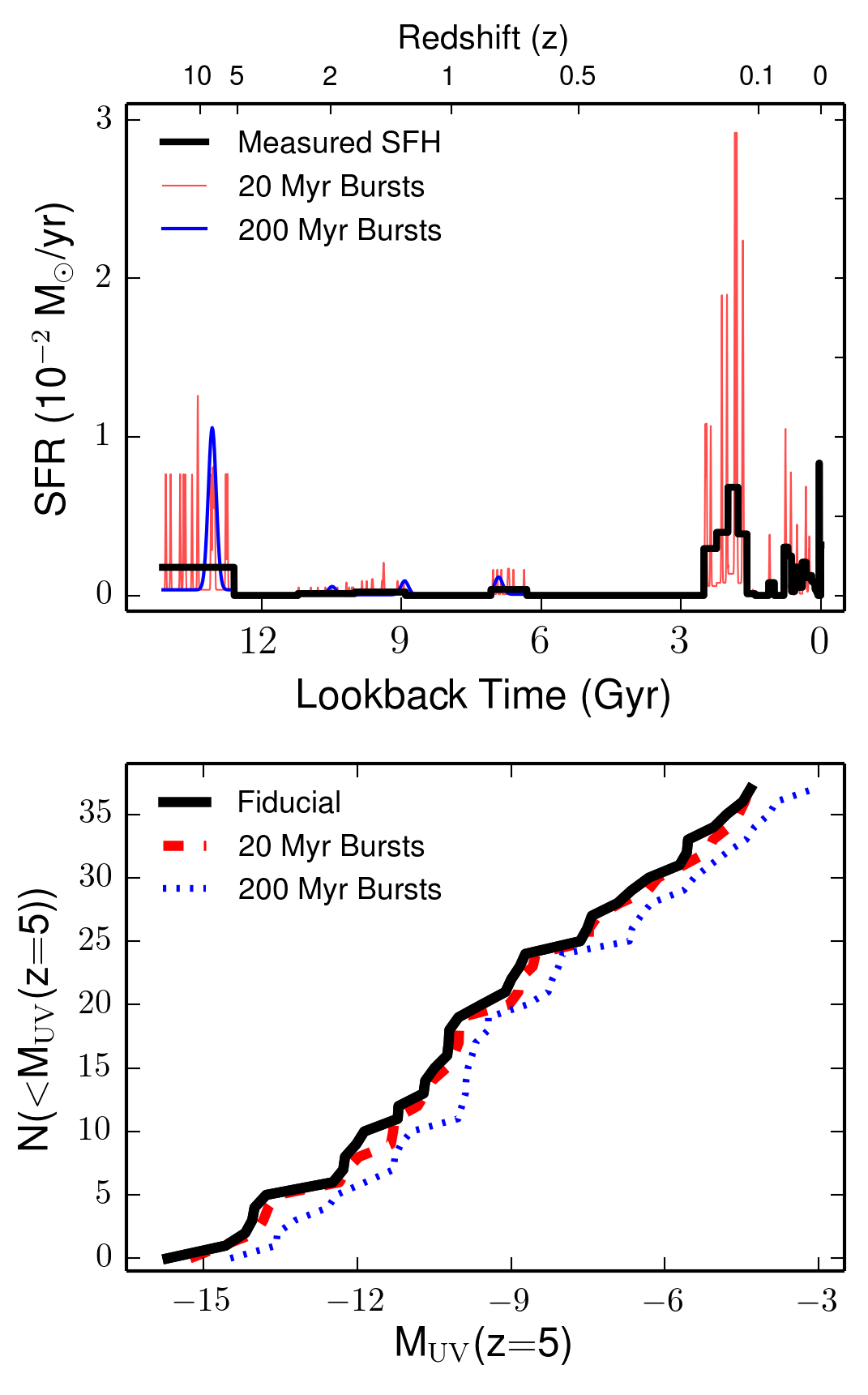} 
\caption{Translation of SFHs into UV LFs. \textit{Top panel:} SFH
  vs. lookback time at the native time resolution (black) for a single
  galaxy. The red and blue lines illustrate the variations in
  intra-time bin bursts that we considered (see \S \ref{sec:sfhs} for
  details).  \textit{Bottom panel:} The cumulative $z=5$ UV LF for all
  37 galaxies in the sample. The thick black line is the UV LF
  assuming the native time resolution of the SFHs. The red and blue
  lines illustrate the variation in the mean UV LF for model SFHs that
  include bursts on 20 and 200 Myr timescales.}
\label{fig:sfhexample}
\end{figure}


\section{Methodology}
\label{sec:methods}

The primary data used in this analysis are the SFHs of 37 LG dwarf
galaxies that were measured by modeling their {\it HST}/WFPC2-based
resolved star CMDs \citep{weisz2014a}.  UV luminosities as a function
of redshift are then constructed by coupling the SFHs to the Flexible
Stellar Population Synthesis code \citep[FSPS;][]{conroy2009,
  conroy2010a}.  Once various sources of incompleteness are included,
we fit a Schechter function \citep{schechter1976} to a combination of
these reconstructed LG LFs and more luminous literature values.  Here, we 
describe this procedure in detail, including the assumptions
and uncertainties in our analysis.

\subsection{Star Formation Histories of Local Group Dwarfs}
\label{sec:sfhs_lgd}

Our sample consists of 37 dwarfs galaxies located within the LG
($\sim$50\% of the known LG dwarf population). Sample galaxies
have low present-day metallicities ($Z\lesssim\,0.2\,Z_{\odot}$),
stellar masses ($10^4\lesssim M_{\star}/\msun\lesssim10^8$), and dust
content \citep[A$_{\rm V}\,\lesssim$0.5;][]{dolphin2003}. Our sample
contains systems of all morphological types and environments (Milky
Way and M31 satellites, isolated systems) found in the LG, including
`ultra-faint' dwarfs. Overall, our sample is broadly representative of
the entire LG dwarf galaxy population.

SFHs of these galaxies are presented in \citet{weisz2014a}. They were
measured by modeling CMDs from archival {\it HST}/WFPC2 imaging using
the maximum likelihood CMD fitting package \texttt{MATCH}
\citep{dolphin2002}. An example SFH derived from CMD fitting is shown
in Figure \ref{fig:sfhexample}. Full details
of \texttt{MATCH} and application to the LG sample can be found in
\citet{dolphin2002} and \citet{weisz2014a}, respectively.  Important
for this study is the fact that the SFH time resolution is, at best,
$\sim10-15$\% of the lookback age. Thus, the highest well-resolved
redshift is $z\sim5$.  This is a fairly stringent limit owing to
lingering uncertainties in stellar evolution and the subtle changes in
isochrones at old ages \citep[e.g.,][]{gallart2005}.

\begin{figure}[t!]
\center
\plotone{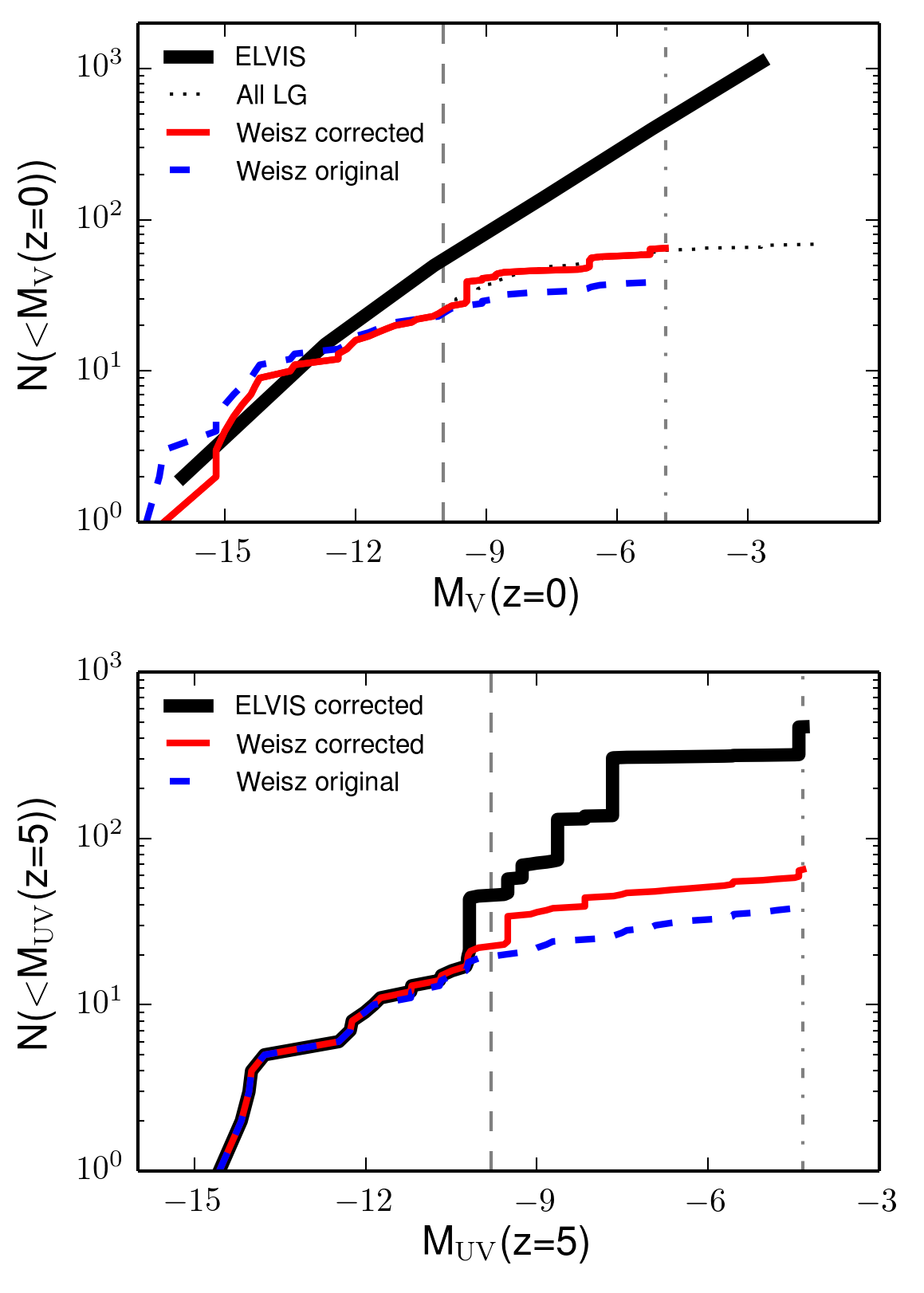}
\caption{Illustration of completeness corrections applied to the dwarf
  galaxy data.  \textit{Top panel:} The $z=$0 V-band luminosity
  function of the original sample (blue), the sample corrected by
  environment (red), the entire LG dwarf LF (black dashed), and the
  predicted LF of the LG from the EVLIS simulation (black solid).  The
  grey dot-dashed line indicates the faint limits of our LG data at
  $z=0$ and the dashed line indicates our conservative sample cut at M$_V(z=0)=-10$.
  \textit{Bottom panel:} Completeness corrections applied to
  the $z=5$ UV LF.  Throughout the paper we use the ELVIS corrected
  UV LFs.}
\label{fig:weisz_uvlf}
\end{figure}

\subsection{From Star Formation Histories to UV Fluxes}
\label{sec:sfhs}

Following the procedure detailed in \citet{johnson2013}, we use FSPS
to translate each SFH into the GALEX rest-frame FUV flux as a function
of time.  As a first step, we correct for aperture effects inherent to
the observations. The {\it HST} field-of-view typically covers a
modest fraction of most galaxies ($\sim5-50$\%), and thus {\it
  HST}-based SFHs are not necessarily globally representative. We
derive an approximate aperture correction by comparing the synthesized
$z=0$ V-band luminosity with the observed V-band luminosity from
\citet{mcconnachie2012}. We multiply each SFH by the ratio of these
quantities and use the resulting re-scaled SFHs and FSPS to
compute the FUV flux profile.  When synthesizing the UV fluxes with
FSPS, we assume a constant metallicity of $0.2\,Z_{\odot}$ and no
internal dust extinction.  The latter assumption is reasonable as
the UV spectral slopes of faint high redshift galaxies
are consistent with a dust-free spectral energy distribution
\citep[e.g.,][]{Bouwens12b, Dunlap13}. 

For lookback times older than a few Gyr, the native time bins of the
SFHs are larger than the timescale of UV emission
($\sim100$Myr).  We must therefore explore the effects of intra-bin
star formation variation on the synthesized UV LFs. We quantify the
amplitude of this effect by considering many permutations of stochastic 
intra-bin bursts. Two representative examples are shown in
Figure \ref{fig:sfhexample}. In both examples, 80\% of the total
stellar mass is formed in the burst phase (mass is, of course,
conserved relative to the fiducial SFH), and the bursts have the same
contrast amplitude of 20 relative to the `off' burst star formation
rate.  The two models differ in burst timescales of 20 and 200 Myr
(red and blue lines, respectively).  The 20 Myr model is patterned
after SFHs from fully cosmological simulations of dwarf galaxies
\citep[e.g.,][]{governato2014}, while the 200 Myr bursts illustrate
fairly dramatic changes in the UV flux profile, as discussed in
\citet{weisz2012a}.
 
We assume these two burst models for all galaxies in our sample, and
plot the resulting $z=5$ UV LFs in the bottom panel of Figure
\ref{fig:sfhexample}.  For the purposes of this
exploratory study, we largely mitigate the effects of intra-bin bursts
by placing our UV LFs in 3-magnitude wide magnitude bins.  This is
also desirable as our sample only contains 37 galaxies, yet spans 
a wide range in UV luminosities.

We must also consider both random and systematic uncertainties in
the native SFHs themselves.  Random uncertainties are due to the
finite number of stars in a CMD, and generally scale in
amplitude with the sparsity of the observed CMD. Synthesizing the
random uncertainties for all galaxies via a Monte Carlo analysis is
computationally prohibitive.  Instead, we analyze representative SFHs
to place bounds on the random error range. From this exercise we
conservatively adopt a fractional uncertainty of 50\% on the SFH of
each galaxy.

Systematic uncertainties are due to inherent shortcomings with the
stellar models used to measure the SFHs
\citep[e.g.,][]{dolphin2012}. Their amplitude generally scales
inversely as a function of photometric depth, i.e., deeper CMDs
provide more secure leverage on the full SFH of a galaxy relative to a
shallower CMD because they probe more age sensitive features and older
main sequence turnoffs (MSTOs).  Correctly synthesizing these systematic
 uncertainties is beyond the scope of this current exercise but, in principle, 
can be be done in future analyses.

Finally, we use a \citet{kroupa2001} IMF for our analysis, the same
as adopted by high redshift studies.  However, a variable IMF \citep[e.g.,][]{zaritsky2012,geha2013} would 
systematically affect the shape of the UV LF, but modeling this effect is beyond the scope of this paper.

\begin{figure*}[t!]
\center
\plotone{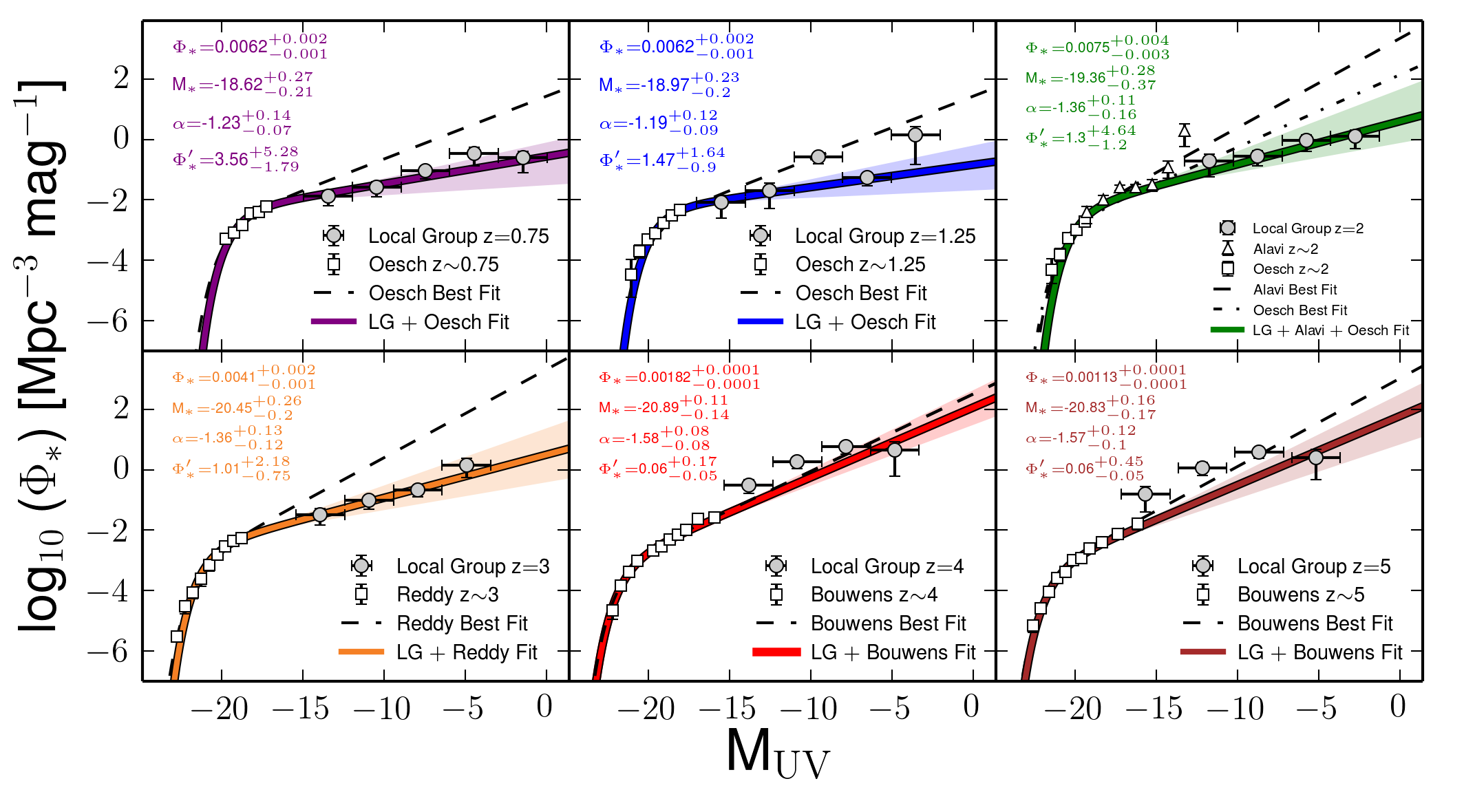}
\caption{UV LFs at redshifts of $z=$0.75, 1.25, 2, 3, 4, and 5. LG data    
  are in grey and select literature values are shown
  as open symbols ($z\sim$0.75, 1.25, 2: \citealt{oesch2010};
  $z\sim$2: \citealt{alavi2014}; $z\sim$3: \citealt{reddy2009};
  $z\sim$4, 5: \citealt{bouwens2014}). The LG data has been normalized
  to the literature data at each redshift for plotting purposes
  only. The thick colored line represent the median Schechter
  function fit to the LG and literature data.  The colored envelope
  enclose the 68\% confidence interval.  Best fit Schechter
  functions from the literature are shown
  as dashed/dot-dashed lines.  Each panel also contains the 16th, 50th,
  and 84th percentiles for the Schechter function parameters.}
\label{fig:uvlfs}
\end{figure*}

\begin{deluxetable}{lcc}
\tablecaption{Rest-Frame UV LFs of Local Group Dwarf Galaxies}
\tablecolumns{3}
\tablehead{
\colhead{\MUV} &
\colhead{} &
\colhead{$\Phi_{\ast}$ (Mpc$^{-3}$ mag$^{-1}$)} 
}

\startdata
\hline
& $z=0.75$ & \\
\hline 
$-$13.44 $\pm$1.5 & & 0.0134 $\pm$ 0.0071 \\
$-$10.44 $\pm$1.5 & & 0.0268 $\pm$ 0.0142 \\
$-$7.44 $\pm$1.5 & & 0.092 $\pm$ 0.0416 \\
$-$4.44 $\pm$1.5 & & 0.3513 $\pm$ 0.2151 \\
$-$1.44 $\pm$1.5 & & 0.2459 $\pm$ 0.165 \\
\hline
& $z=1.25$ & \\
\hline
$-$15.53 $\pm$1.5 & & 0.0083 $\pm$ 0.0059 \\
$-$12.53 $\pm$1.5 & & 0.0207 $\pm$ 0.0155 \\
$-$9.53 $\pm$1.5 & & 0.2653 $\pm$ 0.1063 \\
$-$6.53 $\pm$1.5 & & 0.0539 $\pm$ 0.0244 \\
$-$3.53 $\pm$1.5 & & 1.3885 $\pm$ 1.2372 \\
\hline
& $z=2$ & \\
\hline
$-$11.75 $\pm$1.5 & & 0.191 $\pm$ 0.1342 \\
$-$8.75 $\pm$1.5 & & 0.2834 $\pm$ 0.1503 \\
$-$5.75 $\pm$1.5 & & 0.9242 $\pm$ 0.5239 \\
$-$2.75 $\pm$1.5 & & 1.2877 $\pm$ 0.7885 \\
\hline
& $z=3$ & \\
\hline
$-$13.92 $\pm$1.5 & & 0.0333 $\pm$ 0.0189 \\
$-$10.92 $\pm$1.5 & & 0.0956 $\pm$ 0.0454 \\
$-$7.92 $\pm$1.5 & & 0.2162 $\pm$ 0.0899 \\
$-$4.92 $\pm$1.5 & & 1.4591 $\pm$ 0.8935 \\
\hline
& $z=4$ & \\
\hline
$-$13.82 $\pm$1.5 & & 0.3184 $\pm$ 0.151 \\
$-$10.82 $\pm$1.5 & & 1.8523 $\pm$ 0.7706 \\
$-$7.82 $\pm$1.5 & & 5.9041 $\pm$ 2.6702 \\
$-$4.82 $\pm$1.5 & & 4.5149 $\pm$ 3.91 \\
\hline
& $z=5$ & \\
\hline
$-$15.66 $\pm$1.5 & & 0.1608 $\pm$ 0.1206 \\
$-$12.16 $\pm$1.5 & & 1.1793 $\pm$ 0.5334 \\
$-$8.66 $\pm$1.5 & & 3.8774 $\pm$ 1.454 \\
$-$5.16 $\pm$1.5 & & 2.573 $\pm$ 2.1117 

\label{tab:uvlfdata}
\end{deluxetable}

\subsection{From UV Fluxes to Luminosity Functions}
\label{sec:vmax}

Having constructed UV LFs for our 37 galaxies, we now correct for two
sources of incompleteness.  First, we account for known LG dwarfs that
are not included in our sample. The sample is split by environment
into galaxies associated with the MW, M31, and field, which allows us
to account for the different selection functions (owing to different
galaxy discovery surveys) and different average SFHs across
environments. Within each environment we weight the galaxies in our
sample by the LF of that environment. The result of this correction is
shown in Figure \ref{fig:weisz_uvlf}. Note that our
sample only extends as faint $M_{\rm V}=-4.9$ at $z=0$, and we ignore
any galaxies fainter than this limit.

We next correct for undetected galaxies expected to exist in the LG,
either because of incomplete sky coverage or flux and/or surface
brightness limits.  While a number of empirical estimates of `missing'
galaxies exist for MW satellites \citep[e.g.,][]{koposov2008,
  tollerud2008}, there are no such calculations for the entire
LG. Therefore, we use predictions made by the ELVIS N-body simulation
of LG-like environments \citep{garrisonkimmel2014} to estimate the
number of expected low-mass galaxies located within the LG at $z=0$.
As shown by the solid black line in the top panel of
Figure \ref{fig:weisz_uvlf}, nearly an order of
magnitude more faint systems are expected to exist than are currently
known. In this correction process, we propagate the fractional Poisson 
errors per magnitude bin from our intrinsic $z=0$ V-band LF to the 
UV LFs at all redshifts.

We illustrate the effect of the various volume corrections on the
$z=5$ UV LF in the bottom panel of Figure \ref{fig:weisz_uvlf}. Clearly, 
accounting for missing galaxies is the dominant correction
factor.  

The normalization of the LG LF need not be the same as the high
redshift field galaxy LF, both because of differential evolution of
the LG with respect to the field and because of sample variance due to
large scale structure in both samples.  We address this by including
separate normalization terms when fitting Schechter functions to the
LG and literature LFs.  Specifically, we model the combination of our
LG data and literature data with a Schechter function of the form

\begin{equation}
\Phi(M) = 0.4 \, {\rm ln}10 \, \Phi_{\ast} \, (10^{-0.4({\rm M} - {\rm M}_{\ast})})^{\alpha + 1} \, e^{-10^{0.4({{\rm M}-{\rm M}_{\ast})}}} \, ,
\label{eqn:schechter}
\end{equation}

\noindent where $\Phi_{\ast}$, M$_{\ast}$, and $\alpha$ follow the
standard definitions of normalization, characteristic magnitude, and
faint end power-law slope.  At each redshift we fit the literature data and 
reconstructed LG LFs simultaneously and require M$_{\ast}$ and $\alpha$
to be the same for both Schechter functions; only the normalizations
are allowed to differ between the two datasets.  We then write the
log-likelihood function for the combined datasets as

\begin{equation}
{\rm ln}\, P_{\rm total} = {\rm ln}\, P + {\rm ln}\, P^{\prime} \propto \nonumber
\label{eqn:likelihood}
\end{equation}

\begin{equation}
 -0.5 \, \Big( \, \sum_{i} \frac{(y_{{\rm literature}, i} - \Phi_i)^2}{\sigma_{ {\rm literature}, i}^2}  + \, \sum_{j} \frac{(y_{{\rm LG}, j} - \Phi_j^{\prime})^2}{\sigma_{{\rm LG}, j}^2} \Big)\, ,
\label{eqn:likelihood2}
\end{equation}

\noindent where $P$ and $P^{\prime}$ are Gaussian likelihood
functions, $y$ and $\sigma$ are the UV luminosity values and
uncertainties per magnitude bin from the literature and LG,
respectively, and $\Phi$ and $\Phi'$ are the Schechter functions for
the literature and LG data. We place broad top-hat priors on each of
the four parameters to construct a posterior probability distribution,
$P(\Phi_{\ast}$, M$_{\ast}$, $\alpha$, $\Phi_{\ast}^{\prime}$ $|$
y$_{\rm literature}$, $\sigma_{\rm literature}$, y$_{\rm LG}$,
$\sigma_{\rm LG}$) and sample it using the ensemble affine-invariant
Markov chain Monte Carlo routine \texttt{emcee}
\citep{foremanmackey2013}. At each redshift we first fit only the
literature data in order to verify consistency in recovered parameters
with published values, and then proceeded to analyze the joint
literature and LG datasets.

\begin{figure}[t!]
\center
\plotone{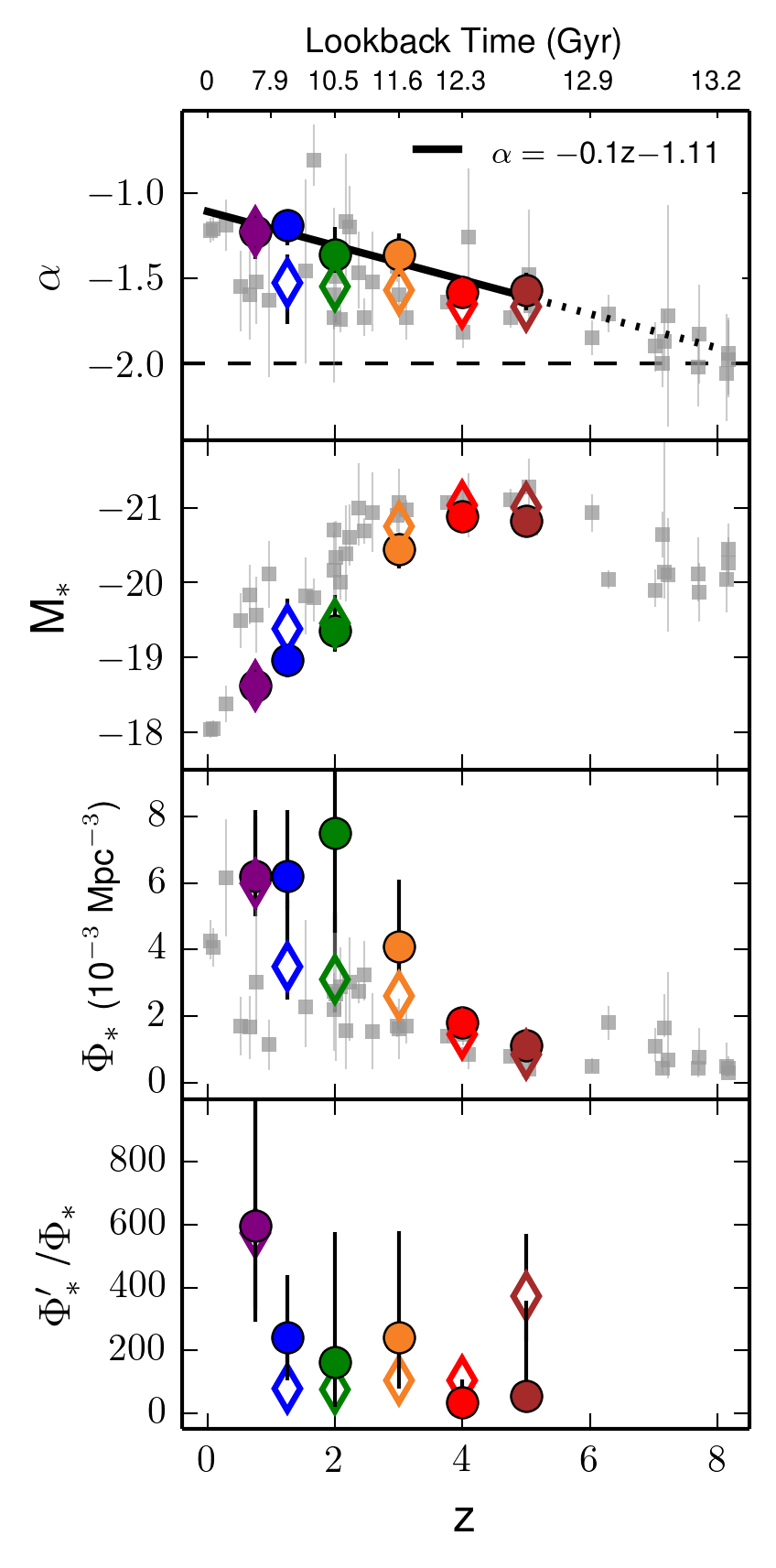}
\caption{Derived Schechter parameters as a function of
  redshift. In each panel, the large circles are measurements from the
  present work, open diamonds are for results when using galaxies with 
  M$_{V}(z=0) \lesssim-10$, and the grey squares are literature data referenced in
  \S \ref{sec:dicsussion}. In the top panel, the solid black line
  reflects the $\alpha-z$ relationship measured from our data and the
  dotted line is an extrapolation to $z=8$.  The bottom panel shows
  the ratio of the normalization parameter for the LG data and the
  direct high redshift measurements.}
\label{fig:params}
\end{figure}


\section{Results \& Discussion}
\label{sec:dicsussion}

The main result of this Letter is shown in Figure \ref{fig:uvlfs},
where we plot UV LFs from $z=0.75$ to $z=5$. Included in each panel
are the reconstructed LFs based on LG dwarfs (grey), literature data
points (open symbols), best fit Schechter functions from the
literature (dashed lines), and the median Schechter function of the
posterior distribution (thick colored lines) along with 68\%
confidence interval (shaded envelopes). As discussed in \S
\ref{sec:sfhs} only statistical uncertainties are included for the LG
data. Our reconstructed UV LFs are listed in Table \ref{tab:uvlfdata}.

There are several important features in this Figure.  First, the the
LG dwarf data allow an estimate of the UV LF to extremely faint
limits. At $z=2$ the fossil record extends the UV LF fainter by a
factor of $\sim10^4$ compared to the lensed observations of
\citet{alavi2014} and $\sim10^6$ fainter than other existing
data. Similar gains are seen at all redshifts, and are particularly
notable at $z=0.75$, where the limits of the LG data extend down to a
remarkable \MUV $\sim-1.5$.  Synthesizing UV fluxes for $z\lesssim0.5$
becomes challenging because of the need to include uncertain
contributions from ancient UV bright populations (e.g., post-AGB and
blue horizontal branch stars).

A second notable feature is the shape of the LG data. Although we have
made several relatively strong assumptions to construct these UV LFs,
they exhibit a shape that is in good agreement with direct
observations of the UV LF at higher luminosities.  Perhaps most
importantly, there is no evidence of a turnover in the UV LFs down to
extremely faint limits --- the data are consistent with a single
power-law slope at faint luminosities over a dynamic range of
$\gtrsim10^4$.

A key advantage of adding in the LG data is increased accuracy and
precision in the recovery of Schechter parameters compared to using
high luminosity data alone.  As is clear from Figure \ref{fig:uvlfs},
the direct measurements \citep{arnouts2005, wyder2005, sawicki2006a,
  yoshida2006, iwata2007, mclure2009, ouchi2009, reddy2009, hathi2010,
  oesch2010, bradley2012, oesch2012, sawicki2012, schenker2013,
  mclure2013, alavi2014, bouwens2014} of the UV LF generally only
probe the bright end and the knee of the LF, which yields a
significant amount of uncertainty in and degeneracy between the
Schechter parameters.  

Our resulting Schechter parameters are shown in
Figure \ref{fig:params}, along with literature values based only on
the direct high redshift measurements.  In general, we find broad
consistency with literature values of $\alpha$ at each redshift,
although we highlight the significant uncertainties ($\sim$0.5 dex) in
most literature values compared to this work.  When comparing the other
Schechter parameters, the agreement between our results and the
literature is less impressive, especially at $z<2$.  In this regime
the direct measurements probe a limited dynamic range that results in
large uncertainties and covariances between the Schechter parameters.
Deeper rest-frame UV data at $0.5<z<1.5$ would be valuable in
clarifying these apparent discrepancies.

In the bottom panel of Figure \ref{fig:params} we show the ratio of
$\Phi_{\ast}^{\prime}$ and $\Phi_\ast$, the Schechter function
normalizations for the LG and high redshift LFs, respectively.  Their
ratio encodes information regarding the relative behavior of the LG
and high redshift LF.  We would expect this ratio to be constant with
redshift if the LG population was representative of the high redshift
field population at each redshift.  If instead the LG dwarfs have SFHs
that systematically differ from the (higher luminosity) high redshift
field population, then this ratio should change with redshift.  This
ratio might also evolve if the present LG population represents only a
small fraction of the high redshift progenitor LG population. The
uncertainties on this ratio are large and are broadly consistent with
constant.  Future improvements both in the high redshift data and the
number of LG SFHs could significantly reduce these errors and would be
valuable for placing the LG in a broader cosmological context.

There are several important assumptions and uncertainties in this
analysis.  As shown in Figure \ref{fig:sfhexample}, the intra-bin SFH
variation introduces modest uncertainties in the reconstructed UV LFs.
This can be properly marginalized over in a more comprehensive
analysis.  A more important concern are systematic errors, which are
typically dominant in the SFHs measured from CMDs that do not include
the ancient MSTOs.

We have assumed that the LG dwarfs are dust-free and have a
single metallicity at higher redshift.  This is a reasonable
assumption given current observations, and it is worth noting that
adding a constant amount of extinction to all the LG galaxies would
only affect $\Phi_{\ast}^{\prime}$; in order to change $\alpha$ one
would require a systematic variation of $A_V$ with dwarf galaxy
luminosity.  

Other limitations in the present study include the corrections
required to estimate LFs from the 37 LG dwarfs.  As shown in Figure
\ref{fig:weisz_uvlf}, the incompleteness corrections at the faintest
luminosities are a factor of $\sim10$ and rely on $N-$body dark matter
simulations (e.g., an assumed SHM relation).  However, the corrections at 
brighter luminosities are quite small.  If we restrict our analysis to luminosities
where the LG data are believed to be highly complete, (M$_{\rm V} \lesssim -10$ at
$z=0$), we recover consistent, but less precise, Schechter parameters 
at most redshifts (open symbols in Figure \ref{fig:params}).

Finally, we expect that the present day LG population
represents only a fraction of the progenitor galaxy population owing to the
disruption of dwarf galaxies over time \citep[i.e., mostly accretion onto the MW/M31, but
potentially dwarf-dwarf mergers as well;][]{deason2014}. 
If this disruption process is weakly dependent on the galaxy luminosity 
then our main results will be unaffected.  
We might expect a weak dependence because the
correlation between dwarf galaxy light and total gravitational mass is
very weak \citep[e.g.,][]{strigari2008}, and it is the latter quantity
that more directly controls the efficiency of tidal stripping and
merging.  It should be possible to estimate these effects directly
with current $N-$body simulations.

These limitations and uncertainties can be addressed and reduced with
upcoming facilities. Wide-field capabilities of the LSST will discover
and better characterize the LG dwarf population, reducing the
magnitude of volume and aperture corrections, and minimizing sample
variance considerations \citep{robertson2010}.  {\it JWST} and WFIRST
will provide unprecedented depth to better characterize the brighter
end of the high redshift LF and reach the oldest MSTOs of the dwarf
galaxy populations in and beyond the LG.  

Properly addressing these limitations 
will ensure that the reconstructed LFs of nearby dwarfs will  
play an important role in characterizing the faint galaxy population at high redshifts.  
Such systems are believed to play a critical part in reionization of the universe, but 
are too faint for direct detection in the foreseeable future.  With our novel technique,
we have demonstrated the existence of a large number of faint systems at high-redshifts.  
Their UV LFs show no hint of a turnover down to very faint limits, and suggest
that current extrapolations of the faint-end slope to higher redshifts are justified.


\acknowledgments 

The authors would like to thank the anonymous referee for 
detailed comments that helped improve the paper.
We would also like thank Julianne Dalcanton for insightful
discussion on the interpretation of $\Phi^{\prime}$, and Mike Boylan-Kolchin and James Bullock 
for general discussion of reionization and the LG. DRW is supported
by NASA through Hubble Fellowship grant HST-HF-51331.01 awarded by the
Space Telescope Science Institute. DRW was also partially supported in
part by the National Science Foundation under Grant No. PHYS-1066293
and the hospitality of the Aspen Center for Physics, and by
Hans-Walter Rix and the MPIA. CC is supported by Packard and Sloan
Foundation Fellowships. This research made extensive use of NASA's
Astrophysics Data System Bibliographic Services.

\end{document}